\documentclass[manuscript,screen]{acmart}
\usepackage{todonotes}
\usepackage[utf8]{inputenc}

\AtBeginDocument{%
  \providecommand\BibTeX{{%
    \normalfont B\kern-0.5em{\scshape i\kern-0.25em b}\kern-0.8em\TeX}}}

\setcopyright{rightsretained}
\copyrightyear{2022}
\acmYear{2022}
\acmDOI{XXXXXXX.XXXXXXX}

\acmConference[]{}{}
\acmPrice{}
\acmISBN{}

\begin{document}

\title{Big data ethics, machine ethics or information ethics? Navigating the maze of applied ethics in IT}

\author{Niina Zuber}
\email{niina.zuber@bidt.digital}
\affiliation{%
\institution{Bavarian Institute for Digital Transformation}
    \city{Munich}
  \country{Germany}
  }

\author{Severin Kacianka}
\email{severin.kacianka@tum.de}
\affiliation{%
  \institution{Technical University of Munich}
    \city{Munich}
  \country{Germany}
}

\author{Jan Gogoll}
\email{jan.gogoll@bidt.digital}
\affiliation{%
\institution{Bavarian Institute for Digital Transformation}
    \city{Munich}
  \country{Germany}
}

\renewcommand{\shortauthors}{Zuber et al.}

\begin{abstract}
Digitalization efforts are rapidly spreading across societies, challenging new and important ethical issues that arise from technological development. Software developers, designers and managerial decision-makers are ever more expected to consider ethical values and conduct normative evaluations when building digital products. Yet, when one looks for guidance in the academic literature one encounters a plethora of branches of applied ethics. Depending on the context of the system that is to be developed, interesting subfields like big data ethics, machine ethics, information ethics, AI ethics or computer ethics (to only name a few) may present themselves.  In this paper we want to offer assistance to any member of a development team by giving a clear and brief introduction into two fields of ethical endeavor (normative ethics and applied ethics), describing how they are related to each other and, finally, provide an ordering of the different branches of applied ethics (big data ethics, machine ethics, information ethics, AI ethics or computer ethics etc.) which have gained traction over the last years. Finally, we discuss an example in the domain of facial recognition software in the domain of medicine to illustrate how this process of normative analysis might be conducted.
\end{abstract}

\maketitle

\section{Introduction}
Digitalization efforts are rapidly spreading across societies, challenging new and important ethical issues that arise from technological development. Software developers, designers and managerial decision-makers are ever more expected to consider ethical values and conduct normative evaluations when building digital products. Yet, when a member of a development team looks for guidance in the academic literature, he or she will likely encounter a plethora of branches of applied ethics. Depending on the context of the system that is to be developed, interesting subfields like big data ethics, machine ethics, information ethics, AI ethics or computer ethics (to only name a few) may present themselves. The sheer number of different branches of applied ethics within the broader field of computer science might cause some bewilderment as to where to look for guidance regarding the context of a specific product and might, ultimately, lead to a feeling of overwhelmingness. Additionally, interested developers might consult introductory books about ethics in which they will find discussions of normative ethics usually in four forms: Deontology, consequentialism, virtue ethics  or contractualist principles. While these four forms of normative ethics are connected and are actively influencing the branches of applied ethics (like machine ethics, AI ethics etc.) it is by no means obvious how this is the case. The literature effectively leaves the interested developer stranded. In this paper we want to offer assistance to any member of a development team by giving a clear and brief introduction into two fields of ethical endeavor (normative ethics and applied ethics), describing how they are related to each other and, finally, provide an ordering of the different branches of applied ethics which have gained traction over the last years. Therefore, the goal of this article is to give developers a starting point to orient themselves regarding what kind of literature should be consulted if they want to inform themselves in their specific context as well as how these branches of applied ethics are related to normative ethics which, in turn, is needed to reach a decision on an issue. Finally, we will end by discussing an example using facial recognition technology in the medical domain and describe how an ethical analysis for this use case might be conducted.
\section{Normativity for IT }
In this section, we want to briefly introduce normative ethics, ethical principles and applied ethics within the context of information technology (in short: IT). We highlight the difficulty of applying ethical principles straightforwardly to the domain of information technology. Normative ethics analyzes normative arguments and ultimately formulates one itself. Essentially, it is a judgmental endeavor. We will then briefly discuss ethical principles and, finally, the domain of applied ethics.

\subsection{Normativity}
Normative ethics is, in general, that branch of (moral) philosophy that discusses the notion of morally right and wrong, i.e. in a moral sense: the question how we ought to act. More broadly, normative demands can be illustrated best by means of real-life practices. Many everyday situations do not require explicit normative deliberation: such as rules of politeness or traffic rules. We expect ourselves and other people to simply act and respond appropriately and according to shared implicit rules~\cite{strawson2008freedom,nida2019structural}. These expectations constitute behavioral contexts, through which single decisions and actions become reasonable~\cite{macintyre2013after,nida2019structural}. Thus, we orient ourselves at shared goals or values. Very often we do not need to ponder upon our behavior~\cite{mead1923scientific,mead1934mind}. Nothing seems to interfere: Neither the demands of others nor our own. We cope with this complexity because we develop attitudes (in classical philosophical terms: virtues) that allow us to behave desirably according to a certain practice. Of course, the questioning of the adequacy of those normative demands is necessary. This critical thinking is at the core of moral thinking~\cite{mead1934mind}. 

For instance, consider the example of “love”. Love is the mutual recognition of two or more people who are in love and behave accordingly. They jointly establish a praxis of love: A system of mutual expectations of caring and loving. Therefore, we understand other people's motives for their actions because we can understand the reasons for their behavior. We comprehend why people hold their hands or why they buy flowers depending on our knowledge of them being in love. In fact, we even consider love to be a sufficient reason for their action. Strictly speaking, we presuppose other people’s integrity~\cite[p.83]{nida2018digitaler}, which means that we assume that their beliefs and actions fit into their life plan. Only if reasons or motives do not correspond to what we expect from two loving people, we start wondering and start asking questions about how this particular event can be thought of within a bigger context - how it  can be integrated into a coherent and integer praxis \citep{knauff2021handbook}.

Thus, the objective of ethical justifications is to systematize normative arguments~\cite[p.9]{nida1998ethische}. Statements about what we ought to do in certain situations are put into an order. We then, for instance, speak of the normative conditions of love meaning those moments that are constituents of our idea of love. Hence, we expect certain actions, expressions, and statements to make our behavior comprehensible as love.

\subsection{ Ethical Principles}
In normative ethics, justifications are often based on ethical principles of reasoning. These ethical principles are aimed at individual actions and are usually simplified into four categories: deontological, consequentialist, virtue ethical or contractualist principles. Those ethical theories combine moral beliefs and subsume a single moral proposition under one common criteria~\cite{nida1998ethische}. For example, consequentialist principles judge actions by their consequences, whereas deontological principles focus on aspects of duty. Virtue ethics focuses on character traits, whereas contractualist principles focus on agreement and consent (actual and hypothetical). This is how normative ethics arranges moral arguments into a moral system. However, none of the ethical principles mentioned can adequately justify all the reasons we find in our daily routine.

We live in a world of pluralism: Social norms, obligations, roles, and equity principles may all constitute good moral reasons. Beyond that, moral reasons may collide and, in fact, they often do. Wherefore, it is of utmost interest to highlight the power of judgment and the capacity to deliberate. Critical thinking, or more precisely moral reasoning, scrutinizes the appropriateness of moral arguments. Coming back to our example from above: Most times we actually do know what we ought to do to demonstrate love. We know what is appropriate and also what kind of behavior counts as misbehavior.  
 
However, especially in newly emerging fields of social interaction, such as those created by (information) technology, uncertainty originates precisely because no moral imperatives have yet been formulated~\cite{vallor2016technology}. Beyond that, relevant normative facts have yet to be determined and neither do well-established desirable practices exist. In this case, we need to reflect upon what we ought to do, since no daily routine lends itself as a prima facie explanation to the observation at hand. Yet, when digital devices are introduced into shared life practices, they can transform the normative constituents of those practices. In these cases we talk about digital transformation: Dating apps transform mutual expectations of dating behavior or online games create new common practices of exchange~\cite{flanagan2005values}. That transformative and performative modification leads to an increased need for ethical orientation for both: producers and users. However, this cannot be addressed by reasoning on individual actions but needs to be discussed in regards to which way of life we judge as desirable.

\subsection{Applied Ethics as Domain Ethics of IT}
The field of applied ethics is often referred to as the field of philosophical ethics that deals with the practical application of moral considerations and principles. In applied ethics the focus is set on real-world scenarios and the right or good actions that have to be taken within these concrete situations as well as the moral and ethical considerations that apply to a very specific moral conundrum. 

These “applications” of applied ethics are usually in the form of utilizing principles of normative ethics in a concrete setting. In the medical domain, for instance, one could adhere to utilitarianism (the most prominent form of consequentialist ethics) to address the question of who should be placed on top of the organ transplantation list. The argument would then be: Because we have a scarcity of supply of organs that exceeds demand we should allocate the organs in such a way that the overall utility is increased, e.g. young over old, otherwise healthy over people with multiple conditions or a history of drug abuse etc. However, it needs to be stressed that even within a certain societal system not all individual actions can be expressed by one single principle alone. To always act according to utilitarian standards would certainly not make you the best doctor, since it may not declare desirable actions that favor emotional support, i.e. it may turn moral behavior into mechanical reasoning and thus dehumanize morality altogether. Therefore doctors owe to put their action in the service of humanity and to always pursue the goal of healing all individuals~\cite{montgomery2018genfer}. 

Regarding information technology it is even more the case that ethical principles and Principlism may not be as supportive for systematizing moral arguments. Mittelstadt's discussion highlights the shortcomings of a Principlism approach to IT~\cite{mittelstadt2019principles}: Professional ethics include common objectives and values that can provide guidance to professionals. For example the value of healing in a medical context stands above other values and must not be questioned. Doctors learn their duties guided by the primary value of healing and to act accordingly to this value. This is not true for software engineers. No common value is fixed and could commonly be regarded as the single all-encompassing value that should guide every form of software engineering because software developers work in various contexts and deploy their artifacts in multiple societal domains and contexts. Moreover, the lack of a shared professional history leads to a lack of practices and institutions that could bind the professionals to collective values. This also refers to missing methods and didactics on how to apply normative knowledge in their development work. Since no institution exists that binds developers to align their daily work to adhere to normative good standards besides technical functionality, no legal and accountability mechanisms are implemented. In contrast, these institutions exist in fields such as medicine or jurisprudence in which practitioners are subjected to a licensing system. Those shortcomings hamper a professional ethos~\cite{mittelstadt2019principles}. Additionally, ethical principles such as the universality test~\cite{apressyan2018kant}, the doctrine of impartiality or the utility maximization cannot justify all observable decisions or actions in a reasonable way~\cite{darwall1983impartial}. Thus, the paradigms that dominate modern ethics are too under-complex to ensure judgments about all types of situations~\cite[p.60]{nida2005menschliche}. Insofar as there is not one single ethical principle that can make all situations and types of actions evaluatively accessible, we need to elaborate on different branches of applied ethics: domain ethics. A pluralistic assumption of good reasons rather reverses the relationship between the universal principle and the particular individual case. Concrete daily observable issues become constitutive components of ethical theory itself. 

Domain specific branches of applied ethics are characterized by that ``[t]hey [...] transcend the subject boundaries in order to do justice [...] to the complexity of moral judgment and action by including the expertise of other disciplines, other professional fields and experiences of action''~\cite[p. 4; translated by the authors]{nida1998ethische}. Henceforth, we will use the term “domain ethics” when we refer to these domain specific branches of applied ethics. In contrast to applied ethics in general, which can be characterized by the use of moral principles that are applied to evaluate one’s action, domain specific ethics are based on social subsystems that concern specific areas of human practice~\cite[p.63]{nida2005menschliche}. Prominent examples are medical ethics or bioethics. Thus, we understand by domain specific applied ethics, an ethical inquiry of ``societal subsystems'' that pursue a specific purpose constitutive of the societal domain, which are characterized by domain-specific (sometimes idiosyncratic) moral problems. Medical ethics, for instance, deals with all issues regarding medical care and the values and principles that play a role in the medical setting. The main value that guides through medical concerns is the value of healing. All other values are subordinated. Since the development and use of digital devices affects many aspects of our lives and lifeworld practices we cannot identify one single value that dominates and is, on its own, able to provide normative orientation. Therefore, the field of ethics of ``technology and engineering'' must be delimited by the subject matter itself and by its use in a social subsystem. Hence, we need to combine an ethics of the digital artifact itself with a domain ethics to identify desirable normative aspects as well as to formulate adequate criteria. 

This distinction between the techno-generic perspective (regarding the digital artifact itself) and the structural perspective (which focuses on the effects of societal subsystems) can be described as follows: The techno-generic perspective results from the very nature of the technology. This perspective primarily addresses the developers with the demand to consider those special, idiosyncratic aspects in the design of their digital artifacts. The structural perspective, on the other hand, can be captured by the description of the role of technology in daily interactions, so called praxis~\cite{macintyre2013after,nida2019structural,knauff2021handbook}, and its effects on social systems. Here we address the potential influence of IT-systems on our mutual expectations and demands. Unsurprisingly, this leads to a division of macro-, meso- and micro-ethics in connection with digital technology and engineering.

It is therefore important that the field of domain ethics regarding information technology needs to be systematized in order to know which topics should be discussed~\cite{Zuber2020}. If we consider that specific ethical domains should be able to systematize arguments concerning this domain in order to constitute normative knowledge, we can understand why a lot of different domain ethics can be found with regard to information technology. Technological constraints limit ethical option spaces, user behavior and concerns as well as producers, but also societal subsystems determine the logic of a domain ethics. For instance, considering the domain ethics of “medical ethics” are significant when it comes to the use of digital technology in the context of medicine besides AI ethics that focuses on algorithms or big data ethics that deals with data issues in this context, e.g. tumor detection. However, there is a danger of getting lost in the many individual facets, which is why a further ideal-typical classification into micro, meso and macro is appropriate. In introducing this ordering, we can highlight the interrelationships between individual domains that are important for IT ethics and can be interpreted as a map to navigate between the individual domain ethics and to clarify their mutual relationship. 

\section{Micro, meso and macro ethics of IT}
In this article we propose an ordering that relies on three analytical levels in which the lower levels are always a part of the respective level above. To put it in mathematical terms the lower levels are subsets of the respective higher levels (sets). These three levels are the micro, meso and macro level. On the level of micro ethics we place these domain specific branches of applied ethics that are specific to a certain domain or technology and are thus smaller in scale. Furthermore, at the meso level all issues of the micro levels have already been reflected back into the context of producer or user concerns: For instance, we can ponder upon issues of big data ethics taking into account techno-generic values as well as structural values without focusing the discussion on the feedback to the producer or user. This will happen at the higher level - the meso level. The macro ethics then deals with the entirety of feedback loops and needs to be able to systematize the bigger picture while avoiding a deep dive into detailed issues. Here, even fundamental issues such as ontological questions can be addressed. 

\begin{figure}[h!]
  \caption{Macro-, Meso- and Micro
}
  \includegraphics[width=0.5\textwidth]{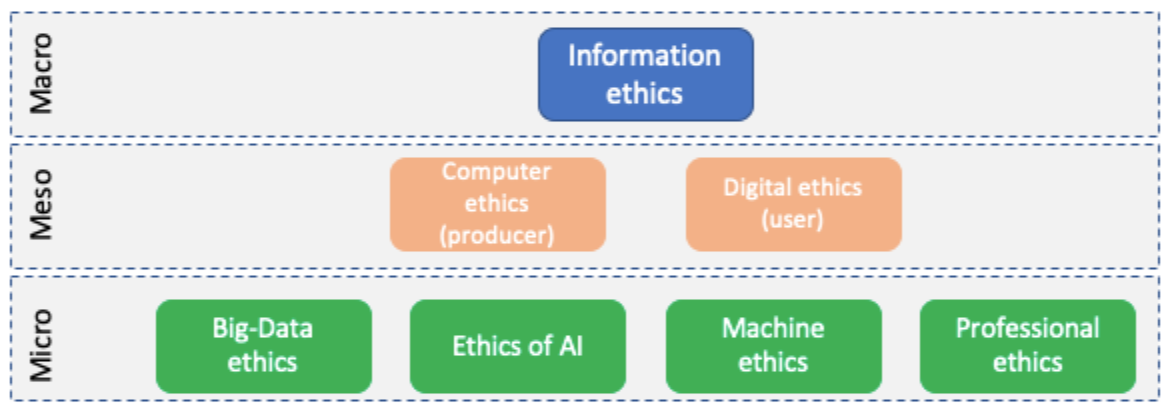}
\end{figure}

In order to find the ethically relevant aspects within the technology itself or within subsocietal fields, it is worthwhile to look at the existing ethical literature on the ethics of IT. However, the dynamics of digital technology is reflected in the sheer infinity of use cases it enables such as process automation, computers, robots, the Internet, digital artifacts, digital products, digital applications, software, software systems, hardware, machines, artificial intelligence and machine learning which are all part of the general notion of digitization or digital transformation. 

Those phenomena are also mirrored in the normative endeavor that theoretically deals with the foundation of these questions: Ethics. There exist a plethora of different ethical approaches that try to tackle the normative and ethical questions regarding IT: Cybernetics~\cite{wiener1950cybernetics}, cyberethics~\cite{tavani1996cyberethics}, digital ethics \citep{capurro2009digital};\cite{ess2013digital,spiekermann2019digitale}, information ethics~\cite{floridi2013ethics}, computer ethics (\citep{johnson1985computer}; \citep{maner1999computer}; \citep{moor1985computer}), internet ethics \citep{langford2000internet}, professional ethics \citep{bynum2004computer}, machine ethics (\citep{anderson2004towards}; \citep{moor1985computer}; \citep{wallach2008moral}), data ethics  \citep{zwitter2014big}, or even Ethics of AI  \citep{bostrom2014ethics}, to name a few. All of these ethical frameworks are concerned with the moral implications of information technology and engineering on individual life plans, intersubjective interactions, or collective goals. 

However, it is important to stress that the various ethical perspectives can be distinguished from one another regarding the object under consideration: For example, data ethics or big data ethics refer to data quality, data processing, and questions of the (proper) application of statistical analysis, while professional ethics is largely concerned with codes of conduct for software engineers or tries to formulate decidedly technical instructions for software developers focusing on value clusters, such as privacy (e.g., privacy by design \citep{schulz2012privacy} or attributions of responsibility (e.g., accountability models \citep{kacianka2021designing}). Machine ethics, on the other hand, is largely concerned with the development of fully autonomous machines, i.e., those that - in the future - might be able to make moral decisions independently and autonomously or at least have the capability to simulate them. Machine ethics shares this question with the ethics of AI in order to discuss the principled ethical basis of a digital moral agency as well as its possible technical design. This is illustrated, for example, by the debate on autonomous vehicles in dilemma situations such as trolley cases \citep{gogoll2017autonomous}\citep{hevelke2015responsibility}. Ethics of AI discusses philosophically encompassing topics such as the possibility of independent thinking ability, the mind-body problem, i.e., whether machines can have a consciousness, or whether machines can have an understanding of language. In addition, the focus of issues in AI ethics is - at least for the moment - on machine learning algorithms as well as data sovereignty and its effects: Non-discrimination, justice as well as data controls are discussed regarding their normative claims.

This approach is skeptically opposed by information ethics, digital ethics and computer ethics. Computer ethics is the oldest of the terms mentioned, if Norbert Wiener's cybernetics is not considered here, it is because he himself did not bother with the ethical dimension of his sociological-structuralist approach of information feedback loops between machines and humans \citep{wiener1988human}. Computer ethics is mainly concerned with ethical issues arising from the use of information technologies and engineering. Thereby, a connection to practical philosophy becomes immediately visible because actions are at the center of ethical reflection. Moor \citep{moor1985computer} expanded the moral question by combining ethical decision making with specifics of information technology and engineering. In doing so, he was able to derive demands from the technology itself. Moreover, bringing these two strands together opens up the possibility of thinking about a responsible development environment. Only if we understand which normative aspects are caused technologically, we can address the developer to reflect ethically while developing software. This approach to ethics of information technology and information engineering is oriented at the perspective of the profession, at least as far as we can speak about a homogenous professionals group of software developers at all. This also distinguishes technology ethics from technology assessment since considerations of the responsibility of software developers cannot be reduced to political evaluation criteria \citep{skorupinski2000technikfolgenabschatzung}. Unsurprisingly, the discussion of the computational foundation of ethics leads directly into educational issues and the question of how to train software engineers \citep{johnson1985computer}\citep{maner1999computer}. Accordingly, computer ethics requires technical knowledge of the object and thus already departs from classical ethics that claims universal validity regardless of their scope of application. 

Digital ethics or digital media ethics, in turn, reflects on which value standards and beliefs are desirable in an interconnected world. In contrast to the above-mentioned ethical approaches that have emerged from software development, the aim here is to examine which ethical theories are particularly suitable for addressing digitization or digital transformation \citep{capurro2020digitale}. Digital ethics thus formulates conditions under which a good life is possible and examines digital artifacts to determine whether or not they are conducive to this goal (telos). This eudaimonistic (as a focus on “well-being” as the highest value) orientation towards formulating criteria that qualify virtues as desirable attitudes as well as discuss their promotion or obstruction by the use of information engineering and technology can be found in \citep{spiekermann2019digitale} and philosophically grounded in \citep{vallor2016technology} and \citep{reijers2020narrative}. Floridi's \citep{floridi2013ethics} comprehensive “The Ethics of Information'' also finds its counterpart in virtue ethics approaches. 

In contrast to computer ethics, digital ethics primarily highlights the moral concerns of the user. These different perspectives, those of the profession, the developers, as well as the end-users, again make both ethics appear as special sub-forms that can be delineated based on their target audience and thus lend themselves as representatives of a higher - meso - level. Capurroemphasizes that one can already speak of digital ethics since the 1940s, even if this type of ethics was called computer ethics at that time. Thus, the traditional branch of professional ethics discussed social effects of information technology \citep{capurro2017digitization}. This already indicates that such sub-ethics cannot be clearly distinguished from one another and are always mutually dependent, insofar as they systematize judgments and lead to normative judgments that are significant for other areas. This is also the reason why we need a macro perspective - information ethics - in order to analyze the big picture of normative inquiry regarding information technology. 

\section{How to navigate the maze: Facial Recognition Systems as an example}
This final section aims to illustrate how a normative analysis might be conducted and as an example, we present the application of facial recognition technology. The concepts from above and their mutual relation will hopefully become more intelligible once their application to a specific case is provided. We will start by discussing facial recognition technology (FRT) in general and then move on to one example of its implementation in the medical domain. Note, we will not provide a full case study but rather an illustration on how the different levels of ethical analysis, as well as the different domains, can guide a solid normative evaluation in a concrete situation.

Before we dig into the concrete example, it is necessary to set the stage by structuring the analysis that is to follow. A good starting point to enable a systematic approach to normative thinking and normative engineering is to distinguish techno-generic considerations from structural considerations (see figure 1 and section 1.3). Each of the two perspectives intends to serve as a systematization of normative arguments that fall under this specific perspective. The consideration of techno-generic aspects aims at understanding and evaluating different engineering methods and techniques that are considered feasible and are consequently discussed in order to investigate how they might contribute to addressing the questions of normativity using features within the technical design. Therefore, these branches of domain ethics (micro ethics) that deal with the technological object itself are of relevance here (section 2). This may also help in examining questions of descriptive responsibility, which means understanding the technological origin of undesirable results. Those microethics are less useful in ordering concepts but to discuss technological constraints or highlight possibilities. 
However, to only address the technical features of a digital artifact is insufficient since the intended deployment of technical systems in social contexts also requires a more precise analysis. This is useful in order to identify relevant normative points of orientation without getting bogged down in an endless collection of seemingly arbitrary and broadly defined values \citep{gogoll2021ethics}. The identification of those normative conditional elements that constitute social domains is just as relevant as a technical consideration. Even if user-centered design approaches already go beyond technical functionality or economic benefit optimization and take aspects of user-friendliness into account, a normative deliberation still has to force further types of thinking. 

\begin{figure*}[h!]
  \caption{Overview of an ethical analysis regarding FRT}
  \includegraphics[width=0.8\textwidth]{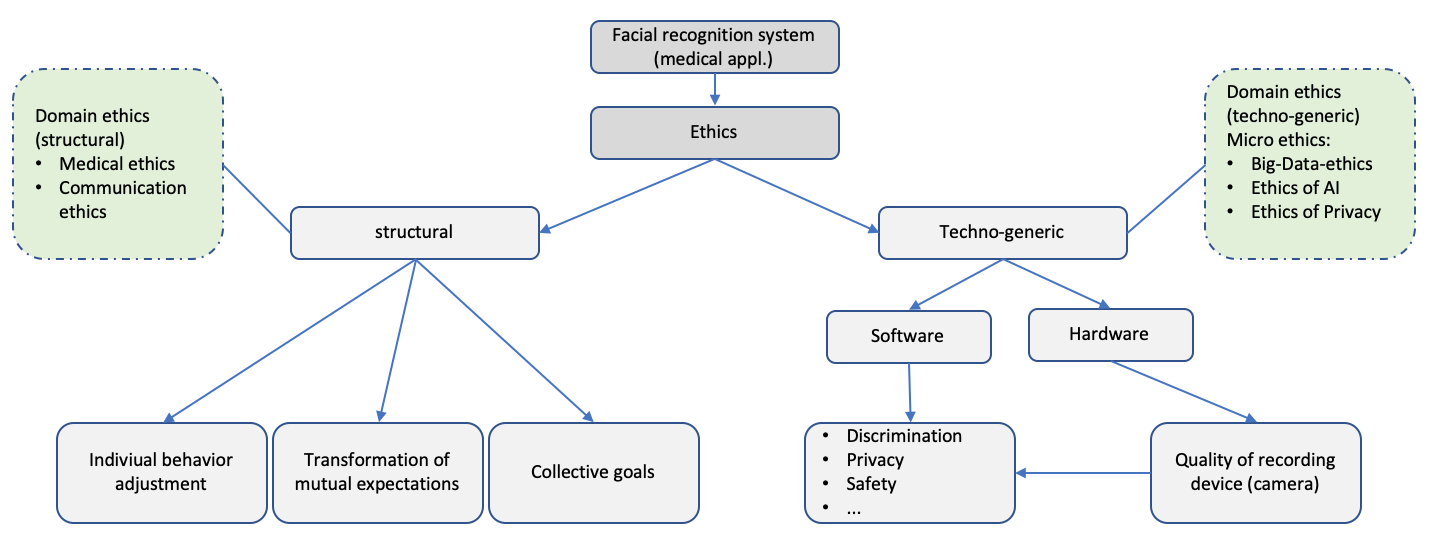}
\end{figure*}

Let us dive into facial recognition systems. In order to make an informed judgment about a reasonable use of said technology, we must consider both perspectives: techno-generic and structural. It is plausible to choose the technical reliability of facial recognition systems to be the first focus of consideration because if a technology malfunctions in a significant way the structural perspective which looks at the impact of society and mutual expectations about the technology will be too chaotic to be addressed adequately. 

As our example we discuss a product using FRT in the medical domain: Face2Gene. Face2Gene is an app that aims to identify rare genetic disorders in people by analyzing people’s faces using FRT and artificial intelligence. It was created by US biotech company Facial Dysmorphology Novel Analysis (FDNA) and focuses on the identification of 10 potential genetic disorders, among them the Noonan syndrome. Their deep-learning algorithm, which they named DeepGestalt, appears to be able to identify the existence of these rare conditions from a person’s photograph alone and seems to achieve a higher accuracy in finding the correct diagnosis compared to their human counterparts, i.e., medical doctors \citep{gurovich2019identifying}. The ultimate goal of the software is to diagnose genetic predisposition early in order to ensure the best personal treatment plan for patients. 

As a first step in our normative analysis we need to avoid undesirable outcomes by achieving technical reliability. Ensuring technical accuracy means to enter a techno-generic perspective. Consequently, the techno-generic perspective needs to establish what is technically meant by facial recognition technology (FRT). Thus, the processes that are digitally replaced must be considered. This approach enables us to localize ethical risks and draws attention to specific inadequacies. All facial recognition systems are technically based on the simulation of face recognition. This means that the software must classify a face in a photo or video as a human face (detection) so that, for example, a person is not accidentally categorized as a gorilla, as was the case with Google Photo in 2015 \citep{grush2015google}. In order to make this technology interesting for surveillance technology or passport control, the assignment of persons in a one-to-many (identification) or one-to-one setting (authentication) is then automated. For this purpose, a metrization or quantification of essential physiognomic features is required, i.e. certain facial peculiarities are transformed into digital data. This process is also known as biometrization. Information is transferred into a form (input), which is transformed into a result (output) that must be interpreted back into information. Further demographic data, such as gender, age or skin color, are not required for these two executions. Face recognition is primarily understood as a system that compares images of faces to determine their similarity using a similarity score \citep{mclaughlin2020critics}. Only then can images or video recordings be compared and matched by using biometric data such as interpupillary distance, forehead width, etc. An increase in the reliability of facial recognition systems can be achieved technologically by adding body posture analysis, which assigns the recordings of the individuals on the basis of further matching parameters. 

The techno-generic perspective further surrounds the discussion about FRT aspects that are related to the recommendation system. This debate was started by Boulamwimi und Gebru in 2018: They highlighted that most systems did not match faces reliably in the wild, especially discriminating against gender and race. Furthermore, they emphasized that biases which were common for machine learning systems were neglected for computer vision systems \citep{raji2020saving}. However, without a data set that is adequately labeled for different skin colors and gender types the algorithm will reproduce undesirable stereotypes. This is because algorithms represent the status-quo insofar as the data set does not represent an “ought-to-be-desirable-state” but displays past reality. Hence, for example, questions about the identification of disease patterns are relevant as they, in turn, convey statements about allegedly normal facial forms.

Chander refers to the intentional, manipulative reproduction of prejudices by automated systems that preserve or even promote the interests of certain groups as a manipulative algorithm \citep{chander2017racist}. On the other hand, viral discrimination is to be distinguished from this, which, unintentionally, picks up on unintentional but existing social distortions and injustices, precisely because algorithms are trained on lifeworld data \citep{chander2017racist}. A viral discrimination must be classified as a far larger risk than intended discrimination since viral discrimination is technically subtle and complicated. This, in turn, leads to focus on transparency and explainability to make the user understand the limits of benchmarks and their realization as well as  application problems \citep{almeida2021ethics}.

Additionally, more data is needed to obtain accuracy \citep{raji2020saving}. However, an increase in  data leads to an increase in privacy issues. These are essentially ethical questions based on a technical origin. Thus, the software developer must be able to take a well-founded position if he or she wants to develop trustworthy software that meets good normative standards. Those tech-generic considerations that are specifically related to algorithms and data can be found in the literature of ethics of AI or big data ethics \citep{arnold2018big}; \citep{floridi2018ai4people}; \citep{morley2021initial}. There, the topics of data protection, the quality of the dataset, security and algorithm bias are discussed. Those ethics primarily address developers to consider those particular idiosyncratic aspects in the design of their digital artifacts. Thus, these micro-ethics are preceded by meta-ethical positions that underlie a pro-ethical stance \citep{floridi2016tolerant}, \citep{turilli2009ethics} or an anticipatory ethics \citep{brey2012anticipating}, \citep{brey2012anticipatory}. Codes of Conducts, regulatory undertakings as well as compliance are increasingly focusing on techno-generic values. Practical life challenges then recede into the background since they are initially located outside technical knowledge and thus, at first glance, have no direct application. 
\begin{table*}[t]
\rowcolors{2}{gray!25}{white}
  \begin{tabular}{p{.1\linewidth}|p{.2\linewidth}|p{.2\linewidth}|p{.2\linewidth}|p{.2\linewidth} }
    \rowcolor{gray!50}
    \textbf{Perspective}&
\textbf{Option Space}&
\textbf{Normative Aspects}&
\textbf{Ethics}&
\textbf{Methods}\\
Techno-generic&
Knowledge of data processing -transmission and storage. Tools and methods of data processing -transfer and -storage&
Technical reliability; idiosyncratic criteria regarding the technology, e.g., bias, privacy...&
Data ethics, ethics of AI, big data ethics, Internet ethics, Computer ethics, Digital ethics, Media ethics&
Value Sensitive Design, Ethics by Design, Models...\\
Structural&
Context of use of the digital artifacts&
Structural analysis of mutual expectations, e.g., doctor-patient relationships, societal trust of data use&
Virtue ethics, narrative ethics of technology; ethics of the data economy.&
Theory of a structural rationality or approaches to the idea of a good life; living conditions (virtue ethics)\\
 \end{tabular}
  \end{table*}
Therefore, it is of importance that ethics can be systematically assigned based on their respective specific object to guide decisions: When developing algorithms, we need to take a closer look at Ethics of AI and ethics of big data to understand techno-generic values. Further, to understand the system’s impact, theories of rationality \citep{nida2019structural}, \citep{knauff2021handbook} or virtue ethics \citep{vallor2016technology} can help to give orientation. Of course, the techno-generic perspective has to be thought of in dependence of normative theories to formulate design features based on its impact. However, the reverse is also true. Thus, ethics that discuss techno-generic normative issues at first support the identification of normatively relevant facts which are of technical origin. A solid theoretical foundation helps to gain epistemic clarity and to identify the essential cornerstones. For this, one has to address the question of which philosophies and which ethical theories are suitable to capture and adequately address both aspects of digital technology. At the same time, a careful distinction must be made between a theory that provides the necessary background knowledge and a methodology that can be used to apply the theory systematically.   The latter are procedures that debate technical implementation possibilities, as it is discussed for example in Ethics by Design, or Value Sensitive Design \citep{dignum2018ethics}, \citep{friedman2019value}. These topics may all belong to the domain of computer ethics and address the developer of AI systems.

The social contextualization of systems requires separate consideration, as the deployment is also relevant regardless of the technology. Once ethical unreliability or unacceptability is argued, non-technical moments leading to undesirable or unacceptable ways of life must also be considered: For example, a technically reliable system would again lead to discrimination if, for example, only rich or white patients will be recommended certain treatment by the recommendation system based on past experience. This again emphasizes that the argumentation of technical reliability has to be distinguished from the discourse of political legitimacy because both debates refer to different normative aspects. Thus, ethical facets that fundamentally affect the software system are to be discussed first, in order to subsequently normatively evaluate structural challenges that arise from the deployment and use of the product. This is because, depending on their application in private or public spaces, facial recognition systems transform normative-structural contexts that can be described as practices (\citep{nida2019structural}, \citep{knauff2021handbook}). For the medical context this implies to ponder upon influences that constitute individual perception, intersubjective expectations or collective goals. Meaning that the machine may intrude on a doctor's relation towards the world, mutual expectations between doctors and patients, such as trust and equal treatment to the best of knowledge and belief, and set collective goals of the practice, namely healing. These normative requirements, which are anchored in the way people live their lives, are given a new dimension by the inclusion of digital applications. And these normative demands, which constitute the respective relationships (recall our example of ``love'' from the beginning), must be taken into account in the design and use of digital applications. 

Any system must serve specific normative desirable moments when integrated, for example, in a medical context. The domain ethics again specifies requirements which have to be considered technically. In the field of medicine, it is thus necessary to subordinate all other values to the values of healing and care and to consider how these can be promoted and not inhibited by any application. This is a virtue ethical approach as to focus on practices that will allow for desirable attitudes. Individually speaking, the doctor has to be able to trust his or her own judgment and needs to be able to make up his or her own mind while using the recommendation system. The doctor needs to understand the limits and the scope of the system and how it may change his or her perception. So they may very well also decide against the use of the application if he or she thinks it appropriate in that case. Intersubjectively speaking, the doctor’s  relationship towards the patients still needs to be based on trust and openness. The doctor must communicate freely with the patient and not be distorted by any tool that does not respect the values that allow for such a communication. Hence, the domain ethics of communication ethics is also to be taken into account. In the field of communication ethics, the normative relevant conditions that enable successful communication are the values of truthfulness, trust and reliability (\citep[p. 59-68]{nida2011optimierungsfalle}). Truthfulness means that we expect that the statements correspond to the actual beliefs of the person. Further, to expect the person to behave truthfully and thus the trust in what they say to be true. We should not only strive to ensure that what we say is truthful, i.e. our own opinions, but also that our expressions are reliable and accurate. Those normative constituents of successful communication must be given priority if we consider successful communication to be a desirable way of living together. Insofar as digital artifacts, and especially FRT systems, undermine or corrupt these values, the design or deployment must be questioned and, if necessary, restructured. These issues are part of digital ethics, that deliberate on the effects of the user individually as well as intersubjectively.

However, in the medical case the value of healing must serve as a point of reference for the technical-generic values, i.e. privacy must be reflected on the relevant value of the social subsystem. This is central to questions of medical ethics: All actions, behaviors or devices implemented here must be guided by the value of healing as the ultimate orientation. Domain ethics such as medical ethics are of importance in order to understand the underlying structure of the social subsystem. If more data means a more accurate recommendation system, and a more accurate recommendation system helps in making better judgments on diagnosis and treatment, then more data might favor the value of healing. Better judgments may include less (unintended) discrimnation by the doctor, faster diagnosis, more accurate diagnosis and in turn targeted and personalized treatment. 

Finally, we bring all systematized arguments of the micro- and meso level together and think about their conditions within the big picture: Which questions need to be discussed politically? Where is the scope of action for software developers? Where can we localize questions of business ethics? Political questions in the context of FRT are, for example, whether we also want to use these technological systems in police work. This requires different domain ethics than in the case of medical use. We cannot decide on those issues individually, only at the developer level. In the medical context, justice issues must be able to be politically debated and justified since the engineer cannot fully control those normative issues by design. It is essential to understand which normative aspects need a thorough argumentation, if reasonable digital artifacts are to be developed.

This will never be an easy undertaking. Therefore, we emphasize that our judgmental power must be constantly exercised in order to be able to navigate this maze. Our hope is that the more normative facets we can map, the better we can reflect on the different aspects and give support to order our thoughts.


\bibliographystyle{ACM-Reference-Format}
\bibliography{references}

\end{document}